# An IoT Based Water-Logging Detection System: A Case Study of Dhaka

Md. Manirul Islam[1], Md. Sadad Mahamud[2], Umme Salsabil[3], A A M Mazharul Amin[4], Samiul Haque Suman[5]

[1] American International University-Bangladesh, Dhaka, Bangladesh
[2] American International University-Bangladesh, Dhaka, Bangladesh
[3] British Columbia Institute of Technology, Burnaby, BC, Canada
[4] American International University-Bangladesh, Dhaka, Bangladesh
[5] American International University-Bangladesh, Dhaka, Bangladesh

`manirul@aiub.edu`

**Abstract.** With a large number of populations, many problems are rising rapidly in Dhaka, the capital city of Bangladesh. Water-logging is one of the major issues among them. Heavy rainfall, lack of awareness and poor maintenance causes bad sewerage system in the city. As a result, water is overflowed on the roads and sometimes it gets mixed with the drinking water. To overcome this problem, this paper realizes the potential of using Internet of Things to combat water-logging in drainage pipes which are used to move wastes as well as rainwater away from the city. The proposed system will continuously monitor real time water level, water flow and gas level inside the drainage pipe. Moreover, all the monitoring data will be stored in the central database for graphical representation and further analysis. In addition to that if any emergency arises in the drainage system, an alert will be sent directly to the nearest maintenance office.

## 1 Introduction

Webster's dictionary defines water-logging as being saturated with water. It occurs when more water flows through an area that it can handle which in turn causes the excess water to become stagnant and remain wherever it is. This phenomenon is seen mostly during the monsoon season. Water-logging is a major problem that many cities face. It is quite common in cities of third world countries, especially those countries that are bestowed with the season of monsoon like Dhaka, Bangladesh. Fig.1 shows the water-logging problem in the city.

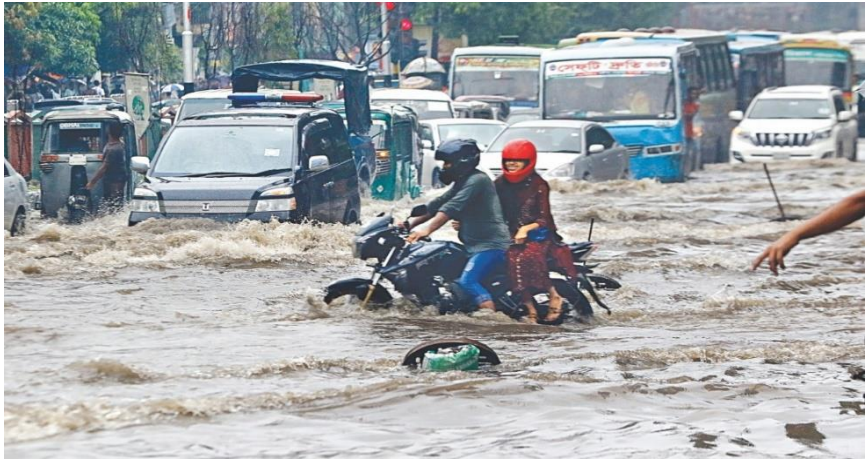

**Fig. 1.** Heavy rain disrupts life in capital Dhaka [1]

It also causes a nuisance the sanitation systems as the sewerage drainage pipes are not that developed to handle large quantities of wastewater at a time. Bangladesh has always been known to be a country with heavy rainfall. The average annual rainfall the country faces is average 200 mm per year. In recent years, Dhaka has faced more rainfall than ever before. This has caused multiple drainage pipes to overflow quite quickly, creating a hazardous situation for its inhabitants.

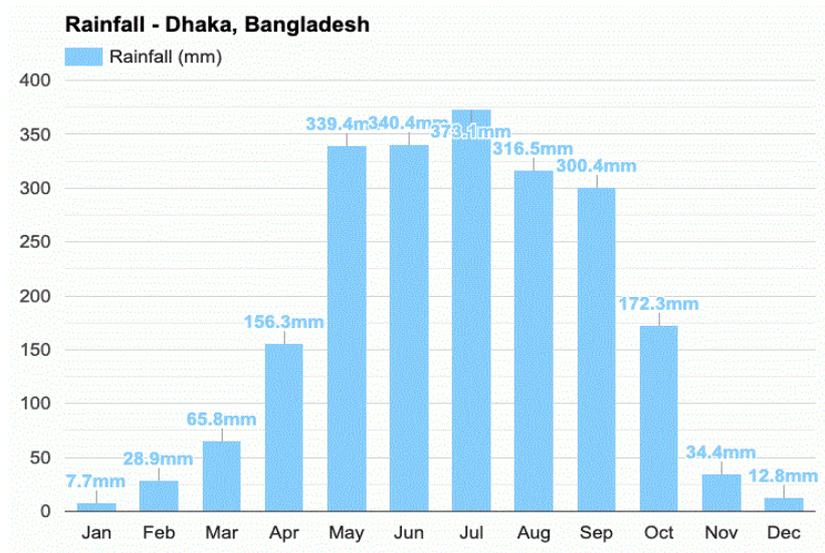

**Fig. 2.** Rainfall in a calendar year [2]

Fig. 2 shows the rainfall rate in a calendar year. Due to this, Dhaka faces massive amounts of water surges through its drains and sewage pipes. This causes the drainage pipes to overflow and the result is multiple waterlogged areas inside a city. Humans are also to blame for this phenomenon known as water-logging. The main cause of water-logging are of course clogged drainpipes, but they are clogged mostly due to human activities it seems. Lack of basic education and sanitation sense has caused people to clog up drainpipes. Criminals are also to blame for water-logging; they are illegally encroaching the waterbodies that serve the city. Lack of proper maintenance causes pumping stations to fail which also aid in the problem of water-logging.

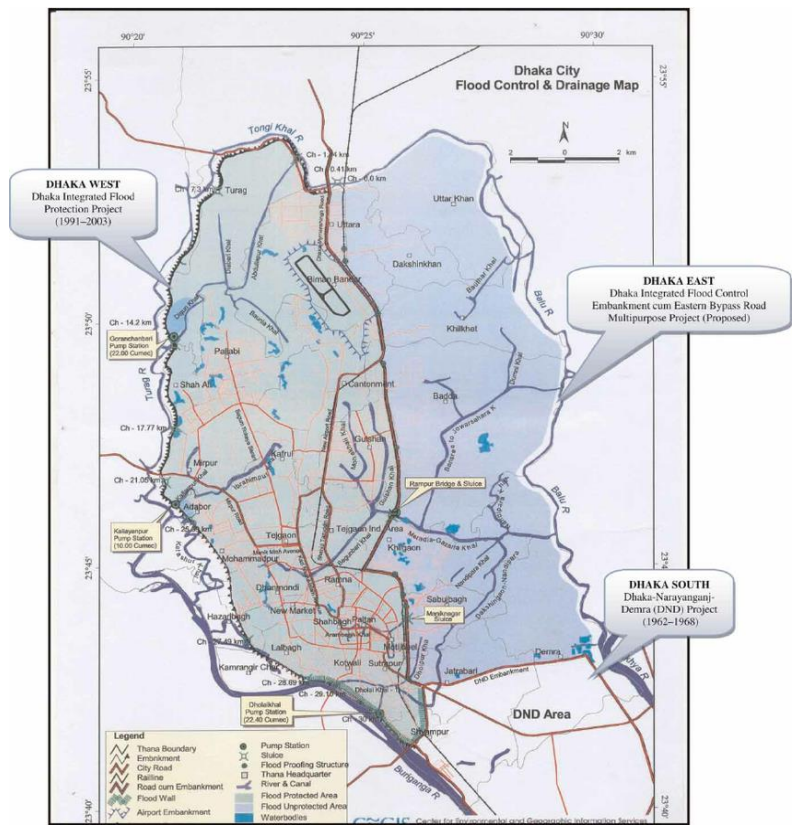

**Fig. 3.** Dhaka city flood control and drainage map [3]

People do not follow the Dhaka Architectural Plan (DAP) when creating new buildings. Fig.3 shows the Dhaka city flood control and drainage map. This sort of unplanned urbanization has also increased chances of water-logging. One important factor is the lack of resources to monitor the city's drainage system properly. All of this could have been avoided had there been proper coordination between the agencies that are supposed to be in control it all. The aim of this proposed system is to be able to detect whether the drainage pipes carrying this excess water out of the city is clogged or not. It sets out to combine Internet of Things (IoT) along with online databases to combat water-logging in the efficient and cost-effective way possible.

## 2      Related Works

Many researchers are working in this field for solving different city water-logging issues. K.L Keung, C.K.M. Lee, K. K. H. Ng and C.K. Yeung developed a Smart City Application for Real-time Urban Drainage Monitoring and Analysis using IoT Sensor which is a case study of Hong Kong. The authors present an IoT model with Artificial Neural Network for predictive maintenance solution for Hong Kong draining system [4]. C. Chen and Y. Pang presented a machine learning based techniques for smart drainage system [5]. V.S. Velladurai with his co-authors provided a garbage alerting system with human safety for smart city [6]. R. Girisrinivaas and V. Parthipan proposed a system called DOMS which is a drainage overflow monitoring system using IoT. This is a case study for India sewage system [7]. M. N. Napiah researched on a flood alert system with android application which is a case study for Malaysia [8].

## 3      Design Architecture

The proposed system is designed using a four-layer architectural standpoint. Fig. 4 shows the architecture block model of the proposed system.

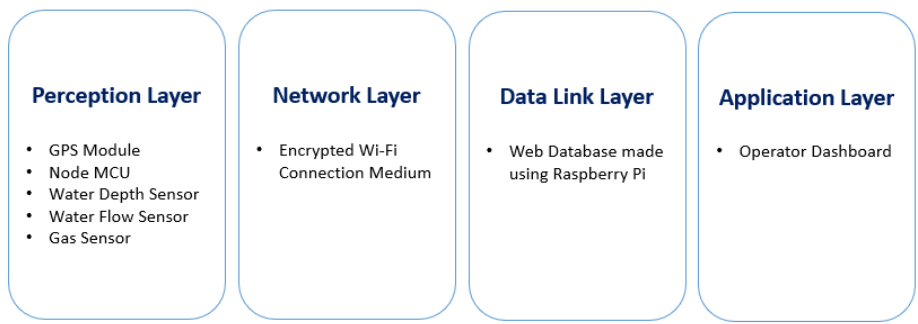

**Fig. 4.** Design Architecture of the Proposed System

The first layer is the Perception layer. This layer is responsible for sensing and reporting. An ESP8266 Node MCU is used here as a microcontroller to control and receive input from the sensors. In this prototype model three types of sensors are used i.e. A Gas sensor, a water depth sensor and a water flow sensor. The sensors are aptly named. They do exactly as they are named. Also, a high powerful GPS model has used for real time location tracking.

Next important layer is Network layer. It is important to secure this layer so that secure communications can take place between the nodes and the server. A wireless communication medium has been used as this eases up the need for kilometers upon kilometers of fussy Ethernet wires. The communication medium is encrypted using A grade security. This layer takes data from the perception layer, turns it into packets of data and sends it over a Wide Area Network (WAN) to the server situated at a fixed location.

Then, the Data Link layer. The server that is part of this layer encompasses an online database within it. The data, sent by the network layer, is interpreted and stored inside the database. The database has various tables for the various sensors that are connected to it. These tables are then used by the next layer, which participates with the actual interaction with the operator.

Lastly, the application layer. This houses the operator dashboard. The operator dashboard takes the data from the database and shows it to the operator. The dashboard is responsible for warning the operator of any possible water-logging as well as various information, such as flow rate, water height and gas level. It also shows the operator the points where the sensors are located on a map. All these operations, from perception layer all the way to the application layer, is occurring in real time.

## 4    PROTOTYPE MODEL

The simulation model of the hardware prototype is given in Fig.5.

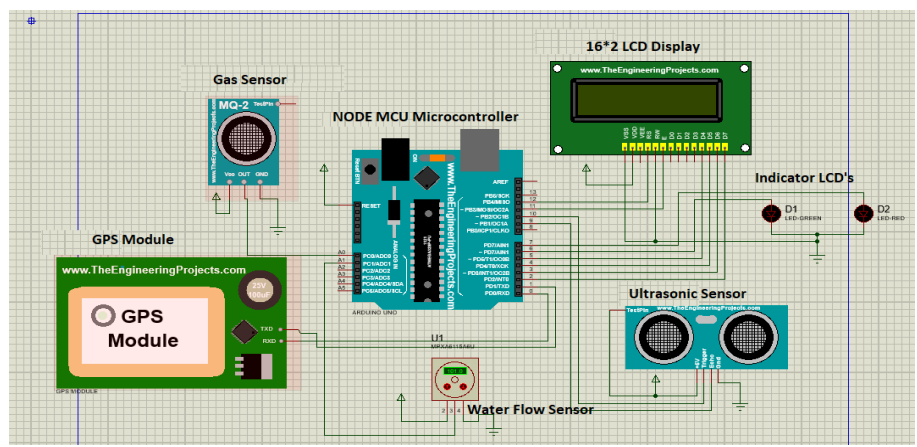

**Fig. 5.** Hardware Prototype Simulation Model

The simulation is done in Proteus Professional simulator [9]. As described in the design architecture, the controlling unit is the Node MCU microcontroller and the various sensors are responsible for collecting all the data. At the end, the calculations are done inside the operator dashboard program. The algorithm used by the program to detect problems inside the drainpipe is given below.

Pseudo code for the warning system:

```
While (application is turned on)
    If (flow rate < setlimit and depth > water level)
      Turn sensor information dock RED
      Turn point on map RED
    Else
      Information dock stays normal
      Point on map stays GREEN
```

Equation to find level of clogging

$$G = PH - D \qquad (1)$$

Where,
  G=Garbage Level
  PH=Height of Pipe used
  D=Depth Reported by the Sensor

As from the algorithm, a warning is issued if a problem does occur. A short description of the hardware used for this system is given below.

### 4.1 Node MCU Microcontroller

For prototyping this system an ESP8266 Node MCU microcontroller module has been used. This microcontroller act both the controlling and network connecting unit. For this system Node MCU is the default gateway for all the sensors to process their data into the server. It is also responsible for fetch all the data from the sensors and execute the defined controlling operations. ESP8266 integrates GPIO, PWM, IIC, 1-Wire and ADC all in one board [10].

### 4.2   GPS Module

The u-blox NEO-6M GPS engine has been used for real time position tracking system. This model is a very high sensitivity for indoor and outdoor applications. The module also works well with a low DC input in the 3.3- to 5-V range which makes this system very much low power consuming. The GPS modules are based on the u-blox NEO-6M GPS engine [11].

### 4.3   Water Flow Sensor

For measuring the water flow rate in the drain pipe a water flow sensor has been used. Fig.6 is showing a water flow sensor model.

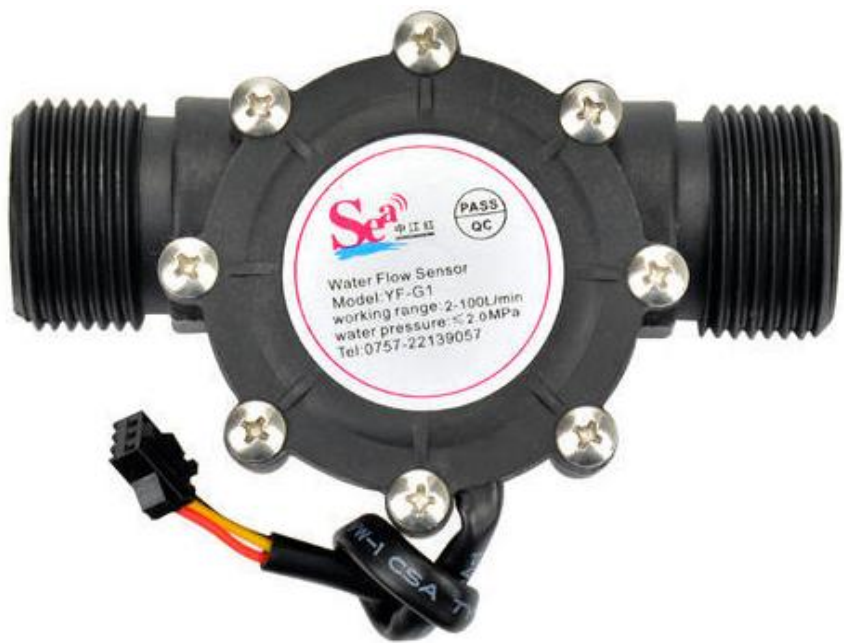

**Fig. 6.** Water Flow Sensor

Water flow sensor consists of a plastic valve from which water can pass. A water rotor along with a hall effect sensor is present the sense and measure the water flow. When water flows through the valve it rotates the rotor. By this, the change can be observed in the speed of the motor. This change is calculated as output as a pulse signal by the hall effect sensor. Thus, the rate of flow of water can be measured [12].

### 4.4 Ultrasonic Sensor

An Ultrasonic sensor has been used for measuring the water depth. It's calculates the length of the drain pipe. If the pipe is full of unused wastage, then the water depth will be high.

$$L = 0.5 * T * C \qquad (2)$$

Equitation (2) is used for calculating the pipe length. Where L is the length, T is the time between the emission and reception, and C is the sonic speed. The value is multiplied by 0.5 because T is the time for go-and-return distance [13].

### 4.5 Gas Sensor

For detecting the gas level inside the drain pipe different gas sensors can be used. For this system a MQ 135 gas sensor has been used. This Gas sensor is suitable for detecting NH3, NOx, alcohol, Benzene, smoke, CO2 [14]. Also, we have used a led indicator for hardware result indication and a 16*2 LCD display for real time hardware result analysis.

## 5 Result Analysis

Fig.7 and Fig.8 shows the simulation test results of the proposed prototype model.

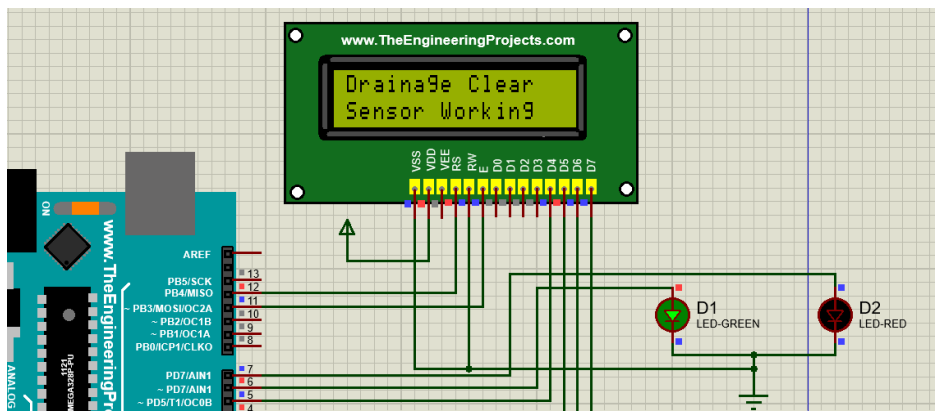

**Fig. 7.** Simulation Test Result at Normal Condition

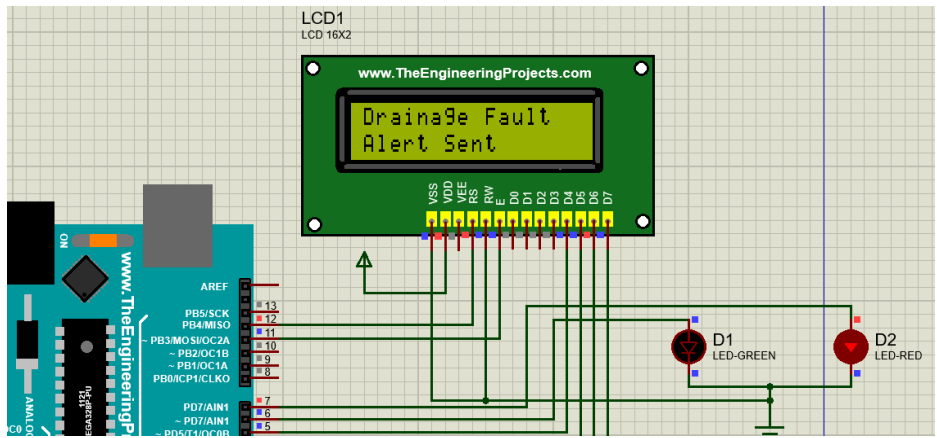

**Fig. 8.** Simulation Test Result at Fault Condition

In normal condition no alert is being shown in the display but if any sensor crossed its pre-defined threshold value it immediately gives an alert to the system.

Fig.9 shows the Operator Dashboard during normal condition. This dashboard has two views; one is sensor window view, and another is map window view. This is what it looks like when everything is operating smoothly and there is no water-logging detected.

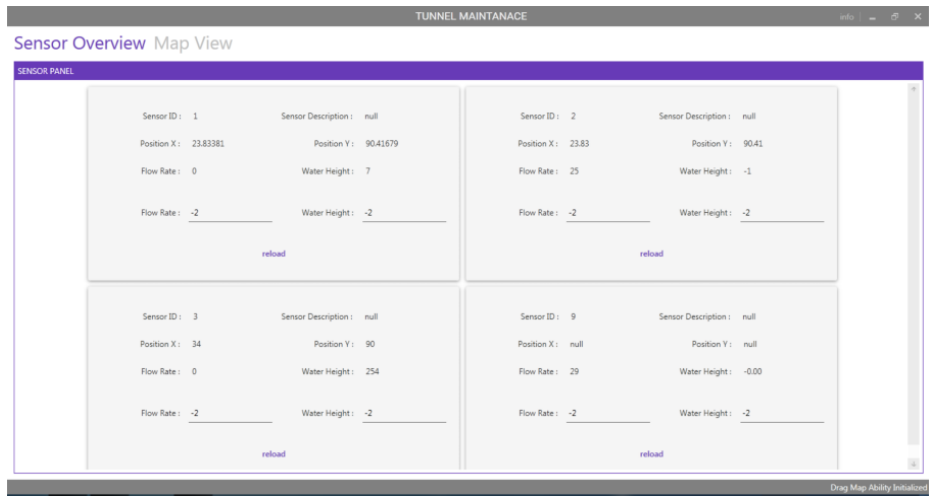

**Fig. 9.** Operator Dashboard during Normal Operation

If water-logging is detected by the system, a warning is being issued and the sensor information panel turns red. Fig.10 shows the Operator Dashboard if a warning is detected.

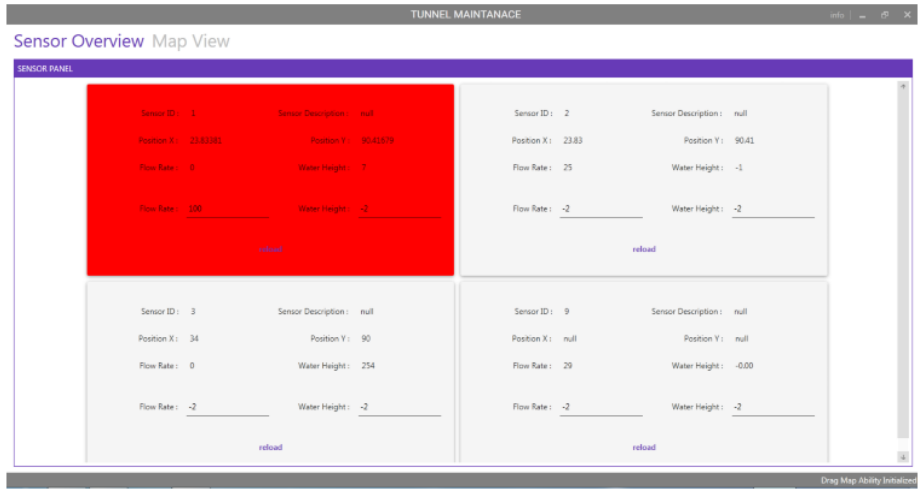

**Fig. 10.** Operator Dashboard when a Warning is being Issued

Fig. 11 shows the map of the area where the sensors are deployed. Since, a problem was detected, a point was marked red on the map. This indicates that the problem lies in this particular area.

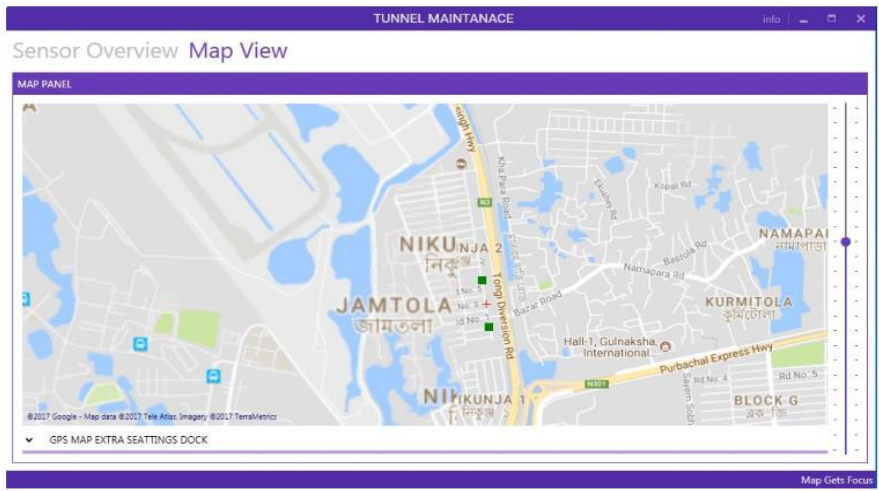

**Fig. 11.** Operator Dashboard showing the Points on the Map

## 6  Conclusion

As water-logging is a serious issue in most of the major cities in the third-world countries crippling the life of the inhabitants in many ways, a simple and low-cost solution is a crying need for effective management of the drainage system. This project of this project was to develop a small yet effective solution for the most densely populated city in the world which can be replicated to other cities of the third-world counties as well. Though the drainage management system alone can't solve this major issue, citizen awareness regarding waste management and abiding the local laws are too important.

## Acknowledgement

Authors would like to express gratitude to Dhaka North City Corporation (DNCC) for their valuable comments, suggestions and support till the very end.